\documentclass[a4paper,fleqn]{cas-dc}

\usepackage[authoryear]{natbib}

\def\tsc#1{\csdef{#1}{\textsc{\lowercase{#1}}\xspace}}
\tsc{WGM}
\tsc{QE}
\tsc{EP}
\tsc{PMS}
\tsc{BEC}
\tsc{DE}

\usepackage{caption}
\usepackage{subcaption} 
\begin{document}

\let\WriteBookmarks\relax
\def\floatpagepagefraction{1}
\def\textpagefraction{.001}
\shorttitle{}
\shortauthors{}

\title [mode = title]{Breaking the Clusters: Uniformity-Optimization for Text-Based Sequential Recommendation}    
\author[1,2]{Wuhan Chen}[%
    ]
\ead{wh.Chen@cqu.edu.cn}

\affiliation[1]{organization={Key Laboratory of Dependable Service Computing in Cyber Physical Society (Chongqing University), Ministry of Education},
                city={Chongqing},
                postcode={401331},
                country={China}}

\affiliation[2]{organization={School of Big Data and Software Engineering, Chongqing University},
                city={Chongqing},
                postcode={401331},
                country={China}}

\author[1,2]{Zongwei Wang}[
    ]
\ead{zongwei@cqu.edu.cn}

\author[1,2]{Min Gao}[
    ]
\cormark[1] 
\ead{gaomin@cqu.edu.cn}
\author[3]{Xin Xia}[
    ]
\ead{x.xia@uq.edu.au}

\affiliation[3]{organization={The University of Queensland},
                city={Queensland},
                postcode={4072},
                country={Australia}}
\author[1,2]{Feng Jiang}[
    ]
\ead{jiangfeng@cqu.edu.cn}
\author[1,2]{Junhao Wen}[
    ]
\ead{jhwen@cqu.edu.cn}

\cortext[cor1]{Corresponding author}

\begin{abstract}
Traditional sequential recommendation (SR) methods heavily rely on explicit item IDs to capture user preferences over time. This introduces critical limitations in cold-start scenarios and domain transfer tasks, where unseen items and new contexts often lack established ID mappings. To overcome these limitations, recent studies have shifted towards leveraging text-only information for recommendation, thereby improving model generalization and adaptability across domains. Although promising, text-based SR faces unique difficulties: items' text descriptions often share semantic similarities that lead to clustered item representations, compromising their uniformity, a property essential for promoting diversity and enhancing generalization in recommendation systems.
In this paper, we explore a novel framework to improve the uniformity of item representations in text-based SR. 
Our analysis reveals that items within a sequence exhibit marked semantic similarity, meaning they are closer in representation than items overall, and that this effect is more pronounced for less popular items, which form tighter clusters compared to their more popular counterparts. Based on these findings, we propose UniT, a framework that employs three pairwise item sampling strategies: Unified General Sampling Strategy, Sequence-Driven Sampling Strategy, and Popularity-Driven Sampling Strategy. Each strategy applies varying degrees of repulsion to selectively adjust the distances between item pairs, thereby refining representation uniformity while considering both sequence context and item popularity.
Extensive experiments on multiple real-world datasets demonstrate that our proposed approach consistently outperforms state-of-the-art models, validating the effectiveness of UniT in enhancing both representation uniformity and recommendation accuracy. The source code is available at https://github.com/ccwwhhh/Model-Rec.
\end{abstract}



\maketitle

\section{Introduction}
\small
 In recent years, sequential recommendation (SR) (\cite{wang2019sequential,hidasi2018recurrent}) has gained widespread applications and achieved considerable success in various domains, including e-commerce, music, news and advertising. As a principal subfield of recommender systems (\cite{su2009survey,zhang2019deep}), sequential recommendation primarily focuses on extracting user preferences from their historical item interactions, offering users appropriate items from the ever-growing number of options.
Despite the considerable achievements of SR, the field has long been trapped with challenges stemming from an ID (identity)-based recommendation paradigm. As illustrated in Fig 1 (a), SR usually represents distinct users and items with unique IDs. These IDs are represented with a large embedding table, where the parameters are updated under the guidance of collaborative filtering information. When new items are introduced to the platform or transferred from other platforms, the system lacks interaction data, making it challenging to learn the embedding parameters. This deficiency adversely affects recommendation quality. Although some methods (\cite{wu2017joint}) incorporate item attribute information to boost recommendations, their encoding techniques remain constrained. As a result, they continue to depend on index-based embeddings and thus face persistent cold-start and cross-domain recommendation challenges.

\begin{figure*}

\includegraphics[width=\textwidth]{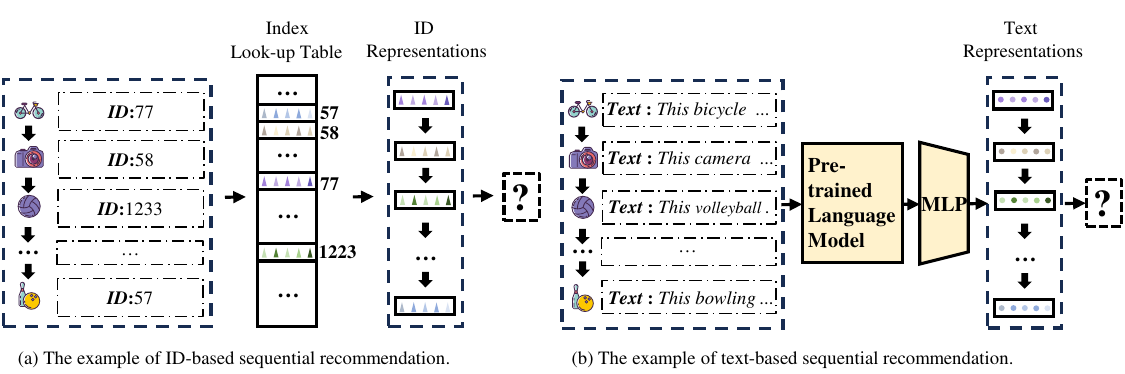}
\caption{\fontfamily{ptm}\selectfont The comparison of ID-based SR and text-based SR.} \label{fig1}
\end{figure*}

As Natural Language Processing (NLP) technologies continue to advance, the original text data can be more accurately represented by the encoder, and the encoder's parameters can be fine-tuned to adapt to downstream recommendation tasks. This development has inspired researchers to investigate the potential of purely text-based SR (\cite{yuan2023go,li2023text}). As illustrated in Fig 1 (b), this approach departs from traditional ID-based SR, which typically regards multimodal information as merely auxiliary. First, text-based SR generates initial representations through a pre-trained encoder, such as BERT or GPT (\cite{radford2019language,brown2020language}) models. Second, the representation vectors are shaped (mapped) to the dimensions required for the recommendation task. By foregoing the reliance on item IDs, this approach effectively mitigates the limitations of ID-based SR and enables a more adaptable and flexible recommendation framework.

Numerous studies (\cite{zhang2024id_embeddings,li2023text}) suggest that text-based recommendation has shown remarkable effectiveness, potentially indicating a new mainstream paradigm in sequential recommendation. Nevertheless, we notice that pure text presents certain drawbacks. Representations generated by pre-trained language models often exhibit high semantic similarity, causing items with similar textual descriptions to cluster more tightly than expected. This clustering effect undermines representation uniformity, a key factor that influences representational quality (\cite{wang2020understanding,wang2022towards,yu2022graph}). Our experiments, as shown in Fig. 2, confirm that text-based SR yields less uniform representation distributions compared to its ID-based counterpart. This observation indicates that improving the uniformity of representations could further enhance the performance of purely text-based SR approaches.

Previous studies (\cite{wang2022towards}) on representation uniformity have focused mainly on ID-based recommendation scenarios, where improving the overall uniformity of the item space was treated as an auxiliary task, typically enforced by introducing penalty factors on inter-item distances. Even when text attributes were introduced, adjustments for uniformity were applied separately to the text modality and the ID-embedding modality. In text-based recommendation, however, certain issues warrant closer attention:

1) Items inherently differ in their semantic relationships. Compared to items outside a given sequence, those within the same sequence should naturally exhibit greater semantic proximity.

2) In purely text-based recommendation, which effectively addresses the cold-start problem, the spatial clustering patterns of popular and less popular items differ from those observed in ID-based systems.

Based on these two points, simply pushing all items equally apart would lead to information loss, and a more nuanced approach is therefore needed.

To address the aforementioned idea, in this paper, we propose a novel framework for Uniforming Text-based Sequential Recommendations (UniT). Specifically, UniT considers the unique characteristics of items and applies varying degrees of “push” to the item representations generated by existing encoders. We design three sampling strategies that achieve a more uniform item representation distribution, i.e. Unified General Sampling Strategy, Sequence-Driven Sampling Strategy, and Popularity-Driven Sampling Strategy. In summary, the main contributions of this paper are as follows:
\begin{itemize} 
\item We observe the non-uniform distribution phenomenon of text-based SR, highlighting an opportunity for improvement in the development of this domain.

\item We believe that the current uniformity strategy should be further refined and propose a unified framework for text-based SR, which aims to yield a more uniform distribution of representations. This framework can be seamlessly applied to most sequential recommendation frameworks.

\item We empirically validate the efficacy of our approach on several real-world datasets, demonstrating its practical applicability and effectiveness.
\end{itemize}

\section{Related works}
\subsection{Sequential Recommendation}Sequential recommendation was initially implemented through Markov chain models (\cite{rendle2010factorizing}), where the current state is dependent on the preceding state, neglecting more distant historical information. With the advent of deep neural networks, researchers utilized Recurrent Neural Networks (RNN) (\cite{lipton2015critical}) for sequence recommendation (\cite{jing2017neural,liu2016context,beutel2018latent}). Hidasi et al. (\cite{hidasi2015session}) proposed GRU4Rec. Another line of work adopts Convolutional Neural Networks (CNN). Yuan et al. (\cite{yuan2019simple}) developed Caser, capturing various lengths of local patterns by sliding convolutional kernels along the sequence. However, both CNN and RNN face challenges in capturing long-term dependencies. Then attention mechanism was utilized for sequence recommendation (\cite{chen2018sequential,kang2018self,sun2019bert4rec}). Compared to the left-to-right unidirectional prediction in the SASRec (\cite{kang2018self}) model, the BERT4Rec (\cite{sun2019bert4rec}) model adopts a bidirectional prediction and masking mechanism. To mitigate the impact of noisy data, FMLP-Rec(\cite{zhou2022filter}) and SLIME4Rec (\cite{du2023contrastive}) combines learnable filters with an all-MLP architecture, allowing the filters to adaptively attenuate noise in the frequency domain. In recent years, some approaches have integrated sequential recommendation with contrastive learning (\cite{zhou2023equivariant,yang2023debiased,zhu2024multi,wang2024unveiling}) or graph neural networks (GNN) to make predictions (\cite{xu2019graph,zhang2022dynamic}). 

\subsection{Text for Recommendation}Initially, text was used as labels in content-based recommendations, grouping objects into a collection according to the labels. During the collaborative filtering phase, text was utilized as supplementary information (\cite{kim2016convolutional}), which can be categorized into two categories (\cite{wu2022survey}). The first is information associated with either users or items. Collaborative Variational Autoencoder (CVAE) (\cite{li2017collaborative}) models the generation of item content while collaboratively extracting the implicit relationships between items and users. The second category is information about a user-item pair, such as review data (\cite{zheng2017joint}) and content like scenario description during interactions. TransNets (\cite{catherine2017transnets}) designed an additional latent module. This module aligns the rating representation with the actual review representations between the item and the user. With the advancement of NLP technology, the capability of text encoders to learn representations has continually increased. A series of recommendation models (\cite{zhang2022dynamic,zhu2019dan,wu2019npa}) based on the Attention mechanism assign attentive weights to text. 

In the era of large language models, textual information is more deeply integrated with recommender systems (\cite{gao2023chat,li2023gpt4rec,friedman2023leveraging}). For example, LLM-TRSR (\cite{zheng2024harnessing}) addresses the issue of large models limiting the text length in sequential recommendations by segmenting the raw textual information of sequences and generating summaries (either through iterative summary generation or chunk-based summary generation).  BinLLM (\cite{zhang2024text}) uses binary encoding to represent the embedding vectors for recommendations, facilitating their integration with text before being input into a large language model. The exploration of interpretable recommendation and multi-task recommendation (\cite{geng2022recommendation}) based on large language models is further pursued.

\section{Non-uniform Distribution of Text-Based Item Representations}
\begin{figure*}
\includegraphics[width=\textwidth]{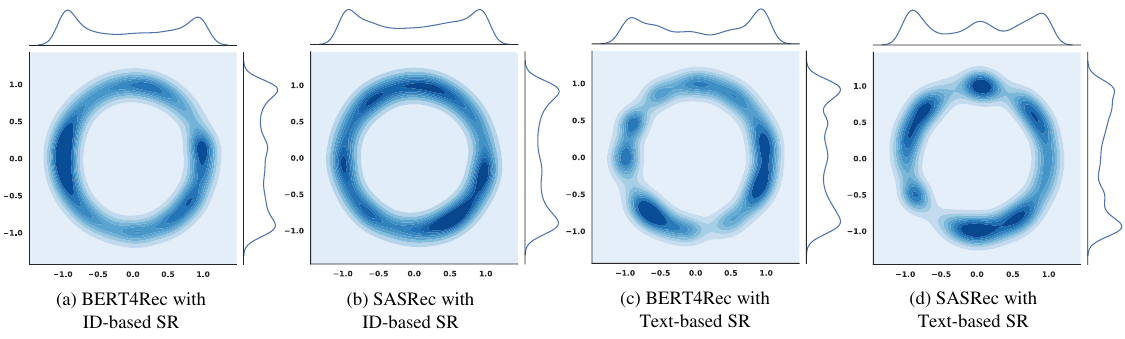}
\caption{\fontfamily{ptm}\selectfont Distribution of item representations learned from the dataset of Amazon-Music.} \label{fig2}
\end{figure*}
\begin{figure*}[!htb]
  \centering
   
  \includegraphics[width=\textwidth]{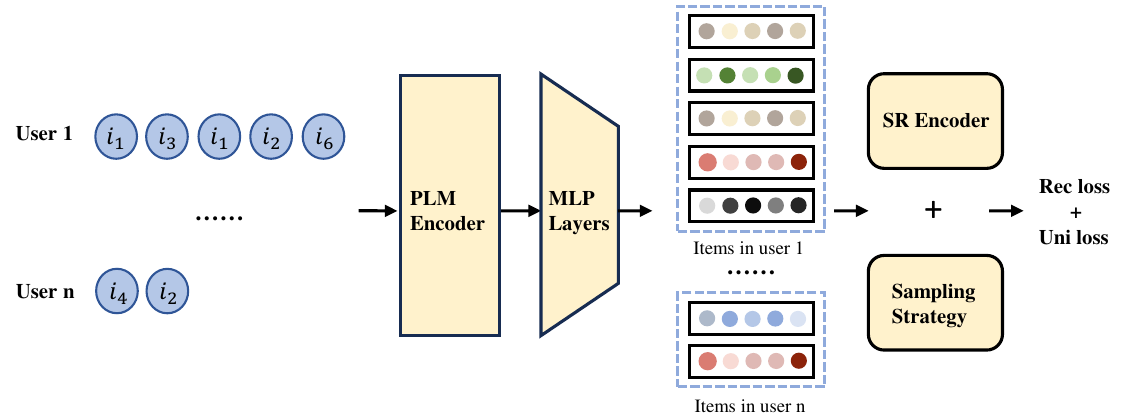}  
  \caption{\fontfamily{ptm}\selectfont An overview of the proposed workflow pipeline for UniT.}
  \label{fig3}
\end{figure*}
In this section, our goal is to determine whether the distribution of item representation in text-based SR achieves uniformity (Without explicit mention, 'text-based SR' in the subsequent sections refers to sequence recommendation using only text). We begin by discussing the inspiration for our research focus. Following this, we provide a comprehensive description of our experimental design. The results of our experiments reveal a noteworthy observation: representations in text-based SR demonstrate a lower degree of uniformity compared to those in ID-based SR.

Drawing inspiration from foundational concepts in the NLP domain, we note that representations produced by text encoders often exhibit anisotropy (\cite{li2020sentence}). In other words, the distribution of these representations varies unequally across different dimensions. One manifestation of this phenomenon is that high-frequency words tend to cluster near the origin of the vector space, whereas low-frequency words are positioned farther away, resulting in a non-uniform representation distribution.
This anisotropy has been documented in commonly used text encoders such as BERT and GPT (\cite{ethayarajh2019contextual,wang2019improving}), which are frequently adopted in recommendation tasks. Our inquiry focuses on whether this anisotropy persists following the embedding mapping process and in downstream recommendation tasks. 

To this end, we conduct a visualization experiment with the Amazon Music dataset. First, we train text-based item representations and ID-based item representations using two sequential recommendation models: SASRec (\cite{kang2018self}) and BERT4Rec (\cite{sun2019bert4rec}). After training, representations are projected onto a hypersphere using t-SNE (\cite{van2008visualizing}). Subsequently, we plot a Kernel Density Estimation (KDE) graph in $\mathbb{R}^2$ visualize the representation distributions (see Fig.~\ref{fig2}). The intensity of the color in specific regions of the plot indicates the degree of clustering in the original high-dimensional space, enabling us to assess the extent of anisotropy in the resulting representations.

Fig. ~\ref{fig2} illustrates the distribution of item representations, where the intensity of the blue color corresponds to the density of these representations. The density of these representations is indicated by the intensity of the blue color. On the left side, the blue shading is relatively uniform across the entire circular area. In contrast, the right side displays a less uniform circular pattern, with darker blue regions indicating areas of higher item concentration. This suggests that numerous items in the original space cluster together more closely on the right side. Consequently, these findings indicate that, compared to ID-based item representations, text-based item representations exhibit a less even distribution.
\section{Method}
To tackle the issue of non-uniformity in text-based sequential recommendation, we present a unified framework, UniT. Fig.~\ref{fig3} illustrates the pipeline of Sequential Recommendation (SR) under UniT. It employs a fixed-parameter text encoder, complemented by several dimension transformation layers, to encode the textual information of sequential items into representations. Subsequently, UniT utilizes a SR encoder, such as SASRec, and applies specialized sampling strategies to these item representations. These two components optimize the representations with both recommendation loss (Rec loss) and uniformity loss (Uni loss). As an example, Rec loss can be instantiated as cross-entropy (CE) (\cite{zhang2018generalized}).
The fundamental principle of the uniformity loss is to increase the representative distances between all items, thereby achieving a more uniform distribution. In subsequent sections, we introduce a range of strategies specifically designed to enhance uniformity. Before presenting these strategies, we first define the necessary symbols and formulas.
\subsection{Problem Definitions}
In the field of sequential recommendation, we have a user set $U$, $u \in U $, and an
item set $I$, $i \in I $.  A user's historical interactions can be represented as  $X_u = \{i_1^u, i_2^u, \ldots, i_k^u \}$ in temporal order, each item can be represented by a latent vector $\mathbf{e}_i \in \mathbb{R}^d$. The task of sequence recommendation is to infer the next item $i_{t+1}^u$ based on \( i_1^u \) to \( i_t^u \) that has interacted. The task is as follows:

\begin{equation}
\underset{\theta}{\mathrm{argmax}} \; P_{\theta}(\mathbf{e}_{t+1} | f(\{\mathbf{e}_1, \mathbf{e}_2, \ldots, \mathbf{e}_t\})),
\end{equation}
where $f$ is a user interest extraction function. In ID-based SR,  $\mathbf{e}_i$ is randomly initialized. In text-based SR it is initialized by a text encoder using the description of $i$.
\begin{figure*}
\includegraphics[width=\textwidth]{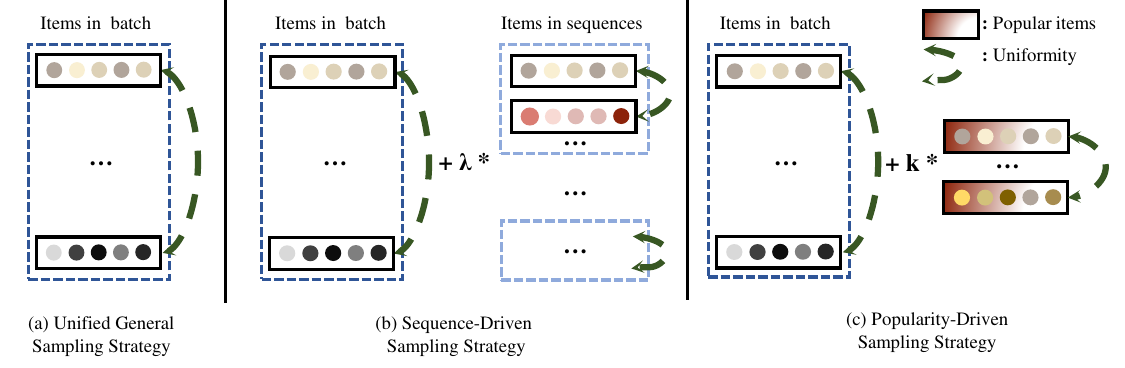}

\caption{\fontfamily{ptm}\selectfont Sampling strategy that assigns different importance to items. Unified General Sampling Strategy\ weakens the uniforming operation within the sequence. Popularity-Driven Sampling Strategy incorporates each item's popularity metric into the calculation.} \label{fig4}
\end{figure*}

\subsection{Unified General Sampling Strategy}\label{A}
The Unified General Sampling Strategy gives equal treatment to all items. We did not employ a two-stage training process, requiring the model to simultaneously satisfy both uniformity and recommendation objectives during a single training phase. In the initial learning phase, since the model has limited information about the personalization of items, the most equitable and straightforward approach is to treat all items equally, assigning them the same uniformity weights. This method ensures that during the model's learning phase, no biases arise favoring particular items. As shown in Fig \ref{fig4} (a), the green curve denotes the repulsion between items. The task is formulated as follows:
\begin{equation}
{L}_\text{general}=log(\frac{2}{||I_b||(||I_b||-1)}\sum_{j=1}^{||I_b||}\sum_{m=j+1}^{||I_b||}\exp(-t(D_{jm}^2 ))),
\end{equation}
where $t \in [0, +\infty)$ is a hyperparameter, and $I_b$ is an item embeddings’ set. Distance $D_{jm}^2$ is as follows: 
\begin{equation}
D_{jm}^2 = \|\mathbf{e}_m - \mathbf{e}_j\|^2,
\end{equation}
$\mathbf{e}_m$ and $\mathbf{e}_j$ denotes the hidden vector of two items $m$, $j$.
In this task, all items in $I_b$ are matched with the other $||I_b||-1$ items to adjust the distances in a pairwise manner, ultimately influencing the global item distribution. For convenience, $I_b$ typically consists of all sequential items within a batch. Since the average sequence length varies across different datasets, the size of the item set $I_b$ may differ in practice, and it can be adjusted through filtering or additional extraction.

\begin{figure}[!t]
\includegraphics[width=\linewidth]{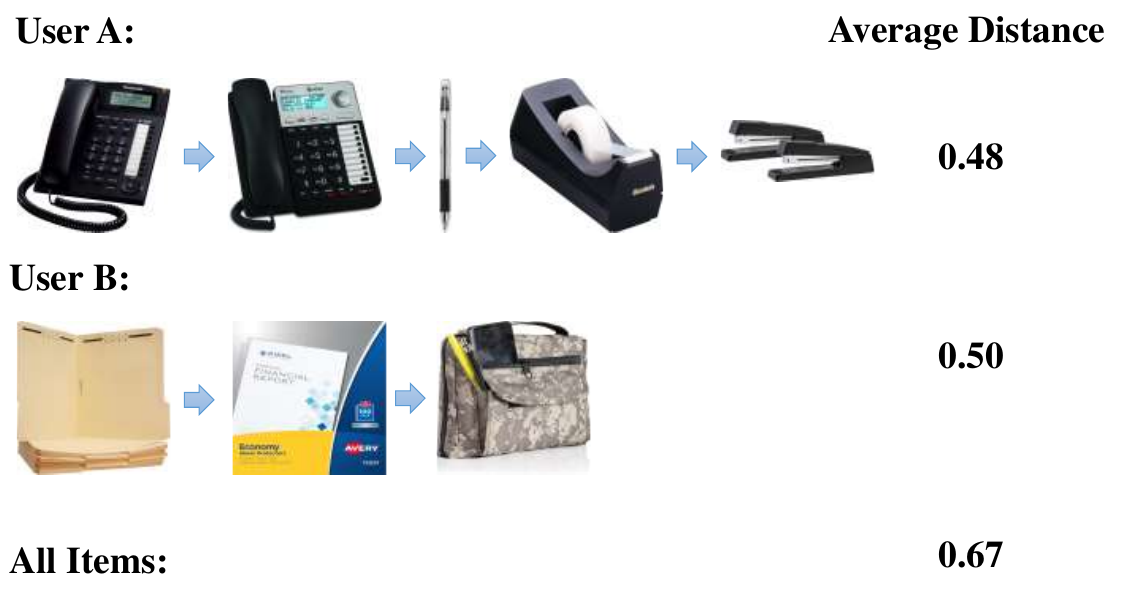}
\caption{The interaction item sequences of two real users.} \label{figcase}
\end{figure}
\subsection{Sequence-Driven Sampling Strategy}
In this section, we discuss the Sequence-Driven Sampling Strategy, as illustrated in Fig \ref{fig4} (b). Though Unified General Sampling is straightforward to implement as it does not necessitate prior information or the specific details of interaction data. However, as we mentioned earlier, all item pairs are actually not equally important. And a user's interests typically remain consistent over a certain duration, implying that two items appearing in the same user sequence always exhibit greater similarity. Fig \ref{figcase} presents a case study. It compares two users from the Office dataset. User A shows a higher interest in telephones, interacting with multiple similar items, while User B prefers paper-related products, with overlapping item attributes in the sequence. To further illustrate it, we calculated the  intra-cluster  distances (average distance between items within sequences) and compared it to the average distance across all items in the Office dataset. 

We found that the average distance within sequences is approximately 86$\%$ of the average distance across all items, the case in  Fig \ref{figcase} also shows the same trend. The results show that items within a single user exhibit a natural tendency to be closer to each other compared to those between different users. Directly distancing such items could lead to inaccuracies in learning both the user's interests and the item representations. Considering this situation, we designed the formula:
\begin{equation}
{L}_\text{sub}=\frac{1}{||U||}log(\frac{2}{(k)(k-1)}\sum_{j=1}^{k}\sum_{m=j+1}^{k}\exp(-t(D_{jm}^2 ))),
\end{equation}
\begin{equation}
{L}_\text{seq}={L}_\text{general}+\text{$\lambda$}\sum_{1}^{||U||}{L}_\text{sub},
\end{equation}
where $k$ is the length of each user sequence which differs in different sequences; $t$ has the same meaning as in Equation (2); $||U||$ is the number of users. Specifically, this method involves traversing each user's sequence and calculating sub-loss functions for the internal item pairs. By iteratively integrating these sub-losses, each scaled by a factor $\lambda$, with the original loss (Unified General Sampling loss), the intensity of uniforming items within the same sequence is effectively diminished. For item pairs across different user sequences, the process remains unchanged. In this method, the probability of two items appearing in the same sequence is inversely proportional to the level of imposed uniformity.

\subsection{Popularity-Driven Sampling Strategy}
Popularity-Driven Sampling Strategy approaches the issue from the perspective of item popularity, as illustrated in Fig \ref{fig4} (c).
In order to explore the impact of the distribution of popular and non-popular items on the uniformity strategy, we first conduct experiments on the distribution of popular and non-popular items in the text space. We select the ML-1M dataset and used SASRec as the training model. To assess an item's popularity, we compute its cumulative occurrence within user interaction data. Then, we categorize the items, defining the top 40$\%$ in terms of occurrence frequency as popular items, and the rest as non-popular items. For each category, we calculate the average Euclidean distance of the item embeddings at optimal performance, resulting in two distances: $\text{distance}_{\text{pop\_text}}$,
$\text{distance}_{\text{cold\_text}}$. We observe that, $\text{distance}_{\text{pop\_text}}$=1.0995, $\text{distance}_{\text{cold\_text}}$=0.8922, meaning that in purely text-based recommendation, non-popular items tend to cluster more closely compared to popular items.
This is problematic because when non-popular items are overly clustered, the system may treat them as similar or identical, preventing it from effectively capturing the distinct value of long-tail items. Consequently, during the uniformity process, it is advantageous to apply stronger uniformity constraints to non-popular items than to popular ones.

Another justification for this approach lies in the presence of both popular and non-popular items, resulting in varying exposure intensities during training. The embeddings of popular items are sufficiently trained, and excessively distancing them from one another may lead to unnecessary information loss. In contrast, non-popular items merit greater emphasis, as improving their representation quality is more beneficial for the system.
In the Popularity-Driven Sampling Strategy,  each item is assigned a popularity score, denoted as $p_{i}$. This popularity metric is then combined with a factor denoted as $t$. The task is formulated as follows:

\begin{equation}
{L}_\text{pop}=log(\frac{2}{||I_b||(||I_b||-1)}\sum_{j=1}^{||I_b||}\sum_{m=j+1}^{||I_b||}\frac{\exp(-t(D_{jm}^2))}{p_{j}p_{m}}).
\end{equation}
Through the formula above, the model’s learning effect on these highly popular item pairs will be reduced, thus preventing their over-optimization.
\subsection{Training Loss}
The overall training loss is as follows:
\begin{equation}
{L}= {L}_\text{Rec}+\gamma{L}_\text{U},         
\end{equation}
where $L_\text{U} \in L_\text{\{general, seq, pop\}}$, and $\gamma \in [0,1]$ is a hyperparameter. Additionally, we maintain its original ${L}_\text{Rec}$. In this paper, we adopt the commonly used cross-entropy loss as the recommendation model, formulated as:
\begin{equation}
{L}_\text{Rec}= - \sum_{u \in U} \sum_{i\in [2,...,k]} \{ \log(\sigma(\hat{y}_{i}^u)) + \log(1 - \sigma(\hat{y}_{j}^u)) \},
\end{equation}
where $i$, $j$ denote positive and negative items, and $y$ means the matching score between representation of $i$ and $j$.
\section{Experiment}

In this section, we conduct experiments to answer three research questions. 

\textbf{RQ1.} Can our UniT outperform the performance of the baselines? 

\textbf{RQ2.} Does the framework truly enhance the uniformity of item representations?

\textbf{RQ3.} How sensitive is the recommendation model to the intensity of the UniT strategies?
\subsection{Experimental Setup}
\subsubsection{Datasets}
We chose (1) "Office Products", "Musical Instruments" datasets from the Amazon Review Dataset (\cite{ni2019justifying})  (2)"MovieLens" from a movie recommendation site . In the following text, these would be written as "Office", "Music", and "ML-1M".  For all, we selected the dense versions. To apply the dataset in text sequence recommendation, we sort the items consumed by a user in chronological order and use the descriptions of the items from the metadata as textual information. The most recent item interaction of each user is used for evaluation. We also remove sequences that are too short. Table~\ref{tab1} provides the details of the dataset after processing.
\begin{table}[width=.9\linewidth,cols=5,pos=h]
\caption{\fontfamily{ptm}\selectfont Detail of datasets. Av. n means the average length of sequences.}
\label{tab1}
\begin{tabular*}{\tblwidth}{@{} lcccc @{} } 
\toprule
Dataset & Users & Items & Av. n & Density \\ 
\midrule
Office & 77,284 & 27,467 & 4.34491 & 1.5819$\times 10^{-4}$ \\ 
Music & 19,513 & 10,445 & 4.52729 & 4.3344$\times 10^{-4}$ \\ 
ML-1M & 6,040 & 3,703 & 44.66805 & 1.2063$\times 10^{-2}$ \\ 
\bottomrule
\end{tabular*}
\end{table}

\subsubsection{Baselines}

Three baselines are used in this paper:

\begin{itemize}
\item{SASRec}
(\cite{kang2018self})
is a personalized recommendation model. It uses stacked unidirectional attention blocks to predict items sequentially.
\item{BERT4Rec} (\cite{sun2019bert4rec}) utilizes the Cloze objective, forecasting a randomly masked item in the sequence by collectively analyzing the item's context both preceding and succeeding it. The number of items to predict is controlled by adjusting the mask rate.
\item{UnisRec} (\cite{hou2022towards}) is originally a framework suitable for pre-training and fine-tuning paradigm. To adapt it to our work, we combined the losses from both pre-training and fine-tuning stages and applied them in a single phase.
\item{LinRec} (\cite{liu2023linrec}) improves transformer-based recommendation models by using an L2-normalized linear attention mechanism instead of the traditional dot-product attention, which helps save time and memory. It can also be integrated with other sequential recommenders. In this paper, LinRec is combined with the prediction mechanism of SASRec, and will be referred to as LinRec hereafter.
\end{itemize}
\subsubsection{Training Details}
For all models, we use a fixed-BERT as a text encoder. Following are 6 MLP layers, which perform transformations with hidden dimensions of [1024, 1024, 512, 256, 128, 64]. The parameters embedding size, max epoch, batch size, and learningRate (using Adam optimizer) were set to 64, 100, 512, and 0.001, respectively. For ML-1M, we set the batchsize to 128. In their respective models, $\gamma$, as well as other hyperparameters of different uniformity loss functions, are adjusted to values that optimize the performance of the recommendation results. A reference value of $\gamma$ is 0.03.

\subsubsection{Evaluation metrics}

HR (Hit Rate) is primarily used to evaluate the recall ability of a recommendation system. The system generates a list of items for each user, and HR indicates whether the recommended list contains the items that the user is genuinely interested in. The formulation is:
\begin{equation}
\text{HR@K} = \frac{\sum_{u \in U} \mathbb{I}(I_u \cap R_u \neq \emptyset)}{|U|},
\end{equation}
where \( U \) is user set, \( I_u \) are target items, \( R_u \) is a top K recommendation list .

NDCG (Normalized Discounted Cumulative Gain) is a common metric for evaluating a recommendation system's ranking performance. It considers both the relevance and position of recommended items, with higher-ranked items having a greater impact on the score. The formulation is:
\begin{equation}
\text{NDCG@K} = \frac{\text{DCG@K}}{\text{IDCG@K}},
\end{equation}
\begin{equation}
\text{DCG@K} = \sum_{i=1}^{K} \frac{2^{rel_i} - 1}{\log_2(i+1)},
\end{equation}
where \( rel_i \) is the relevance score of the item at position \( i \) , IDCG is the DCG calculated based on the ideal ranking.

We adopt HR@(20) and NDCG@(20) as evaluation metrics. 

\begin{table*}[!t]
\captionsetup{width=0.735\linewidth}
\caption{\fontfamily{ptm}\selectfont Experiment results of our model using $L_{general}$ ,$L_{seq}$ and $L_{pop}$. W/O represent models with and without the use of the UniT framework. Bold indicates the performance of sampling strategies surpasses that of the original model.}
\label{tab3}
\centering
\begin{tabular*}{\linewidth}{@{\extracolsep{\fill}}lcccccccccc@{}}
\toprule
\multirow{2}{*}{Dataset}& & \multicolumn{2}{c}{\textbf{SASRec}} & \multicolumn{2}{c}{\textbf{BERT4Rec}} & \multicolumn{2}{c}{\textbf{UnisRec}} & \multicolumn{2}{c}{\textbf{LinRec}} \\
\cmidrule(lr){3-4} \cmidrule(lr){5-6} \cmidrule(lr){7-8} \cmidrule(lr){9-10}
 &  & HR & NDCG & HR & NDCG & HR & NDCG & HR & NDCG \\
\midrule
\multirow{4}{*}{Office} & W/O & 0.03191 & 0.01289 & 0.03444 & 0.01442 & 0.03192 & 0.01296 & 0.02613 & 0.01044 \\
& $L_{general}$ & \textbf{0.03198} & \textbf{0.01323} & \textbf{0.03540} & \textbf{0.01485} & \textbf{0.03196} & \textbf{0.01306} & \textbf{0.02689} & \textbf{0.01090} \\
& $L_{seq}$ & \textbf{0.03447} & \textbf{0.01418} & \textbf{0.03901} & \textbf{0.01646} & \textbf{0.03592} & \textbf{0.01489} & \textbf{0.02664} & \textbf{0.01061} \\
& $L_{pop}$ & \textbf{0.03229} & \textbf{0.01319} & \textbf{0.03588} & \textbf{0.01510} & \textbf{0.03253} & \textbf{0.01340} & \textbf{0.02625} & 0.01043 \\
\midrule
\multirow{4}{*}{Music} & W/O & 0.05275 & 0.02177 & 0.05345 & 0.02180 & 0.05331 & 0.02215 & 0.04881 & 0.01954 \\
& $L_{general}$ & \textbf{0.05298} & \textbf{0.02201} & 0.05345 & \textbf{0.02205} & \textbf{0.05366} & \textbf{0.02222} & \textbf{0.04907} & \textbf{0.02052} \\
& $L_{seq}$ & \textbf{0.05343} & \textbf{0.0218} & \textbf{0.05427} & \textbf{0.02239} & \textbf{0.05411} & \textbf{0.02229} & \textbf{0.05035} & \textbf{0.02053} \\
& $L_{pop}$ & \textbf{0.05406} & \textbf{0.02203} & \textbf{0.05384} & \textbf{0.02195} & \textbf{0.05332} & \textbf{0.02229} & \textbf{0.04933} & \textbf{0.02054} \\
\midrule
\multirow{4}{*}{Movie} & W/O & 0.16545 & 0.06252 & 0.07103 & 0.02562 & 0.16910 & 0.06429 & 0.11010 & 0.04145 \\
& $L_{general}$ & \textbf{0.17301} & \textbf{0.06498} & \textbf{0.07600} & \textbf{0.02750} & 0.16333 & 0.06180 & \textbf{0.11308} & \textbf{0.04227} \\
& $L_{seq}$ & \textbf{0.17075} & \textbf{0.06486} & \textbf{0.07658} & \textbf{0.02719} & \textbf{0.17517} & \textbf{0.06595} & \textbf{0.11292} & \textbf{0.04375} \\
& $L_{pop}$ & \textbf{0.17086} & \textbf{0.06412} & \textbf{0.07889} & \textbf{0.02867} & \textbf{0.17114} & \textbf{0.06467} & \textbf{0.11267} & \textbf{0.04237} \\
\bottomrule
\end{tabular*}
\end{table*}

\subsection{Effectiveness of UniT (RQ1)}

We applied the framework to the aforementioned models and observed the following results:

Table~\ref{tab3} shows most models using the UniT framework outperformed their original performance. Among them, the comprehensive HR performance of the Music, ML-1M, and Office datasets improved by approximately 1.1168\%, 3.8254 \% and 4.1335\%. Accordingly, NDCG performance improved by 1.9796\%, 3.9176\% and 5.1512\%. The degree of performance improvement is related to the characteristics of the dataset itself.

In terms of performance improvement across the three backbone models, BERT4Rec showed the best improvement, followed by SASRec, the improvement of LinRec is similar to that of SASRec. UnisRec being the least effective, which may be related to the use of multiple contrastive losses in the original UnisRec model, potentially conflicting with UniT.

Besides, We want to know the performance comparison between the variant method and the original method. The average performance improvements for $L_{general}$, $L_{seq}$, and $L_{pop}$ are 1.5439\% (HR), 5.0388\% (HR), and 2.4930\% (HR), respectively; and 2.5095\% (NDCG), 5.6016\% (NDCG), and 2.9374\% (NDCG), respectively. This improvement pattern aligns with our previous hypothesis: $L_{seq}$ and $L_{pop}$ outperform the $L_{general}$ approach due to their personalized uniformity settings for users and items. 

The three strategies are suitable for different scenarios. $L_{general}$ is more appropriate when there is a lower proportion of cold-start users and items, as it avoids overfitting by globally optimizing item representations. $L_{seq}$ is suitable for cases where users exhibit high interest persistence, focusing on capturing the semantic relationships within user sequences. $L_{pop}$ is ideal for datasets with a high proportion of long-tail items, as it balances the representation of popular and less popular items, ensuring the inclusion of cold-start items in recommendations.

Furthermore, $L_{seq}$ outperforms $L_{pop}$, possibly because the mapping between item popularity and uniformity is more complex. Future work could consider applying uniformity measures to the relative popularity of items, such as calculating the difference in popularity between pairs of items.

\begin{figure*}[!htb]
  \centering
 
  \subfloat[\fontfamily{ptm}\selectfont Average uniformity variation on the Office dataset.]{
    \includegraphics[width=\textwidth]{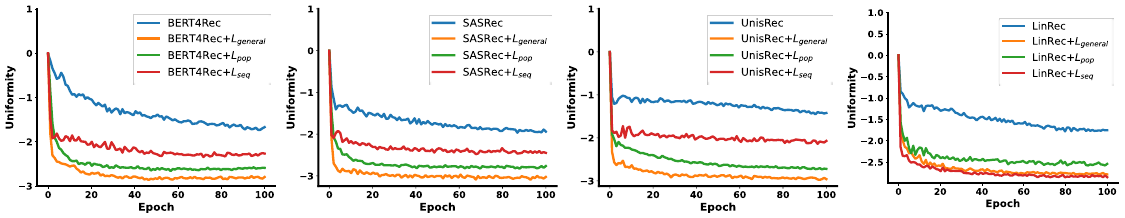}
    \label{Fig.5a}
  }
  \hfill
  \subfloat[\fontfamily{ptm}\selectfont Average uniformity variation on the ML-1M dataset.]{
    \includegraphics[width=\textwidth]{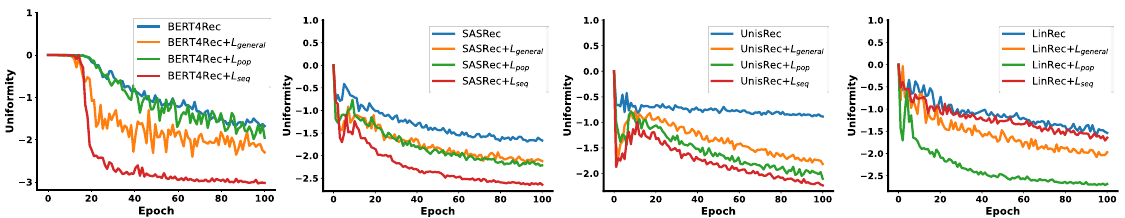}
    \label{Fig.5b}
  }
  
  \caption{\fontfamily{ptm}\selectfont A verification of uniformity on Office and ML-1M datasets. The lower the value, the more uniform a method is.}
  \label{FIG:5}
  \vspace{-2em}
\end{figure*}

\subsection{Uniformity Measurement (RQ2)}
The above experiments demonstrate that our method indeed improves performance. However, one question remains unanswered: whether this improvement is due to the increased uniformity of the representation distribution. This requires a quantitative metric for assessment.
We still use the uniformity loss as the standard for uniformity. However, for fairness, we randomly sampled 2000 items from the representation table each time, treating each item equally and calculating the loss value (as per Equation (2)). This loss does not participate in gradient descent but serves merely as a measure. Here, the value of $\gamma$ is set to achieve the best performance for each method, ranging from 0.01 to 0.05.

The Uniformity of each strategy is shown in Fig.~\ref{FIG:5}. In these graphs, excluding the initial stage and unsmoothed values, the other three lines are all below the blue line, indicating that all our strategies demonstrate better uniformity than the original representations.

We also observed that even a small proportion of  $\gamma$ can significantly improve distribution uniformity, reaching the level of pure ID-based recommendation, demonstrating the efficiency of UniT.

\subsection{Parameter Sensitivity Analysis(RQ3)}
 
In UniT, different strategies are applicable to different loss proportions, so we test the performance stability of these strategies under various $\gamma$ values. For the dataset and model, we choose ML-1M and BERT4Rec. We test $\gamma$ values in \{0.005, 0.01, 0.03, 0.05, 0.1, 0.5, 5\}, and \label{FIG:6}(a) (b) and (c) respectively display the results of ${L}_\text{general}$, ${L}_\text{seq}$ and ${L}_\text{pop}$.

From the figure, the following can be concluded:

1) Under the premise that the absolute value of uniformity exceeds a certain level (varies by model, with a reference value of 2), our framework consistently outperforms the original model at lower values of $\gamma$ ($<$=0.05).

2) The value of $\gamma$ is proportional to the absolute value of uniformity; however, as uniformity continues to increase, the model performance declines, falling below that of the original model. This is because a strong uniformity effect overlooks the individuality of items and the relationships between items, which inevitably affects recommendation performance.
\begin{figure*}[htbp]
    \centering

    \subfloat[$L_{\text{general}}$]{
        \includegraphics[width=0.31\textwidth]{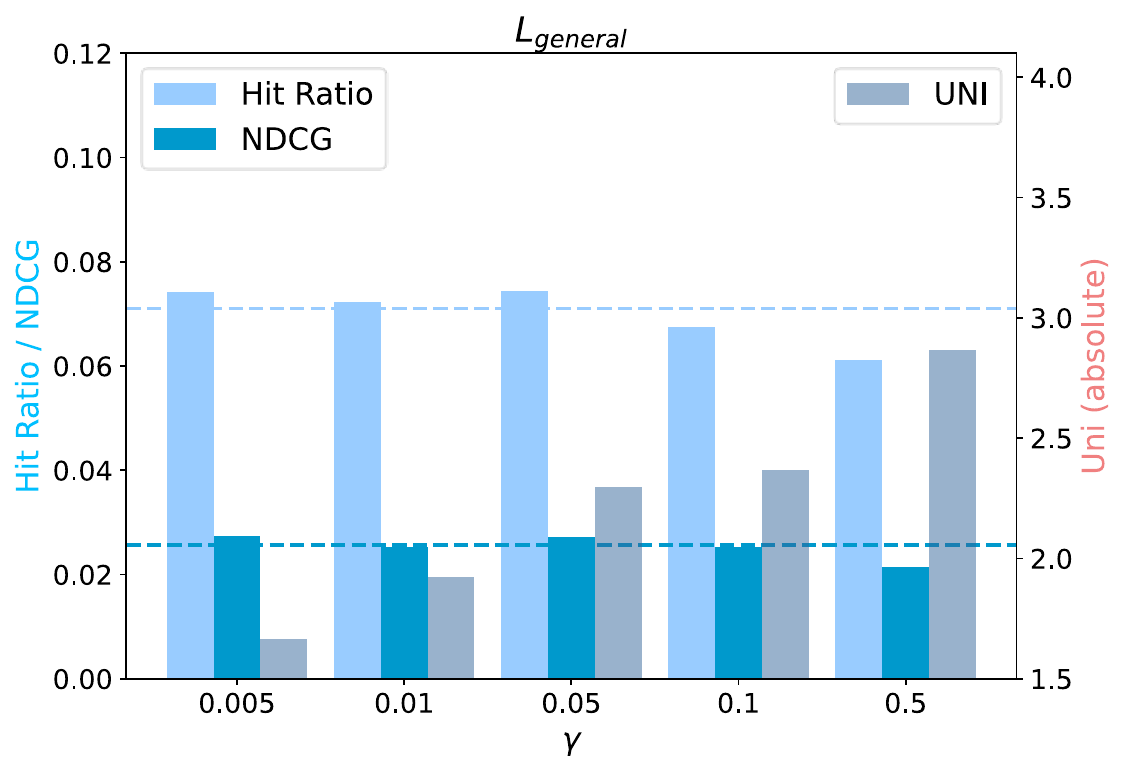}  
        \label{fig:6a}
    }
    \hfill
    \subfloat[$L_{\text{seq}}$]{
        \includegraphics[width=0.31\textwidth]{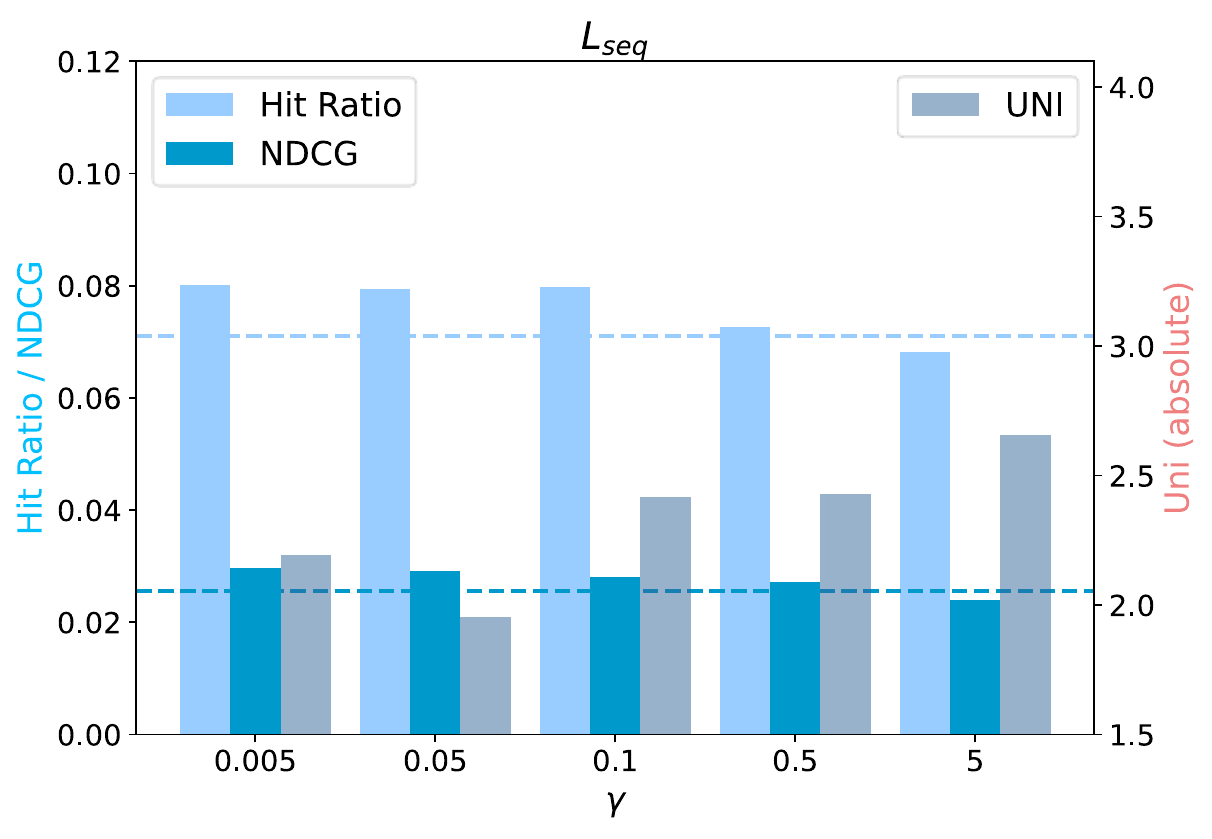} 
        \label{fig:6b}
    }
    \hfill
    \subfloat[$L_{\text{pop}}$]{
        \includegraphics[width=0.31\textwidth]{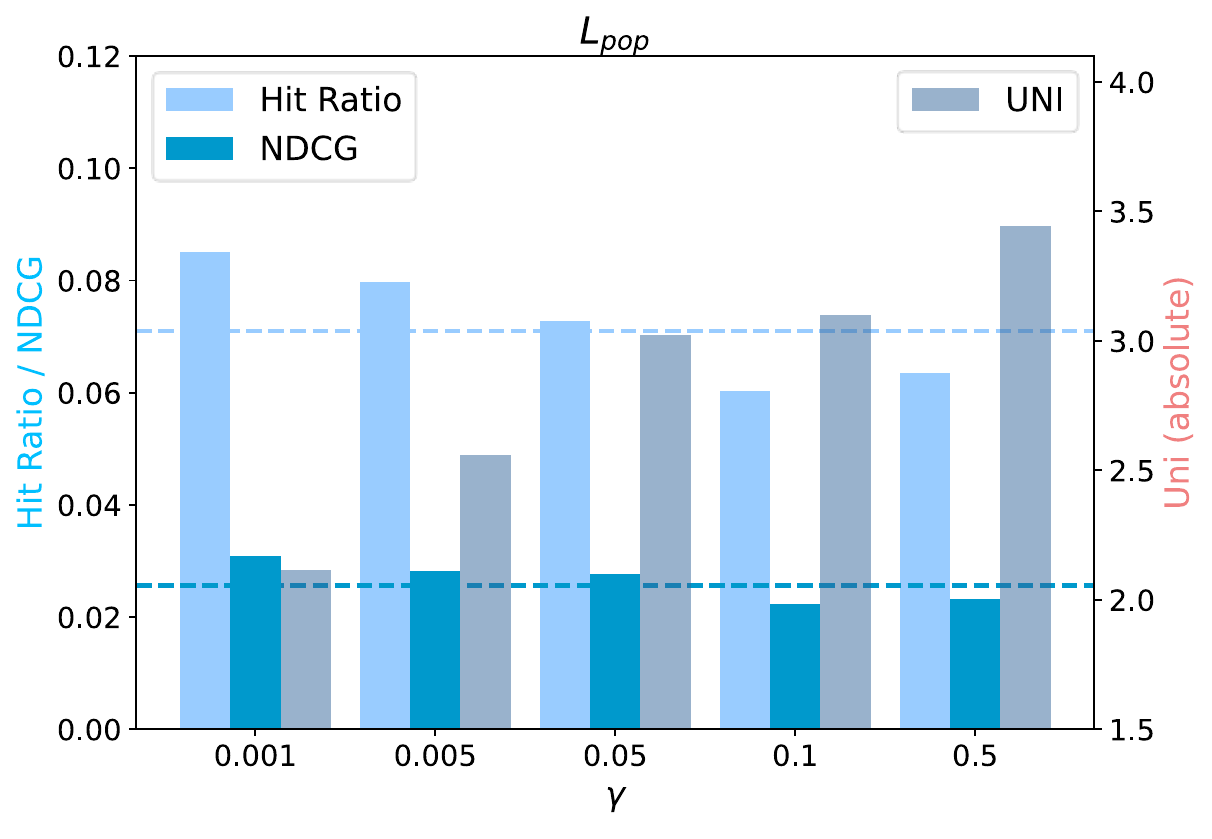}  
        \label{fig:6c}
    }
    
    \caption{\fontfamily{ptm}\selectfont Influence of $\gamma$. The horizontal lines represent the performance of models without using the UniT framework, and the dashed lines' colors match those of the HR and NDCG bars.}
    \label{FIG:6}
\end{figure*}

\section{Conclusion}
In this paper, we find that compared with ID-based item representations, text-based item representations have a more uneven distribution, and then we point out the need to correct this uneven distribution through additional methods. We propose UniT, a framework for pure text sequential recommendation. Experiments demonstrate the effectiveness of UniT and confirm its ability to uniform the overall item representation distribution, which helps to debias in the scenario of recommendation. We introduce three strategies to implement UniT and compare the performance of these strategies. This paper represents a more in-depth study of pure text recommendation, a promising paradigm.

Future research efforts can focus on the following aspects:

Investigating more methods for aligning the textual space with the recommendation space to increase the level of uniformity. In this paper, during the transition from the textual space to the recommendation space, a shift in vector representation occurred, with varying degrees of shift for different items. Future studies could explore the specific factors affecting the degree of shift and uniformity.

Injecting information related to user interests during the text encoding stage. Since raw text attributes, as natural language, may not fully represent the content needed for recommendation, similar item texts might have completely different meanings for users. Special attention should be paid to this during encoding.

\section*{Acknowledgements}
This work was supported by the Science and Technology Innovation Key R\&D Program of Chongqing, China (CSTB2022TIAD-STX0006), the National Natural Science Foundation of China (62176028), and the National Key Research and Development Program of China (2024YFC3014905).

\bibliographystyle{cas-model2-names}

\bibliography{ref}

\begin{thebibliography}{52}
\expandafter\ifx\csname natexlab\endcsname\relax\def\natexlab#1{#1}\fi
\providecommand{\url}[1]{\texttt{#1}}
\providecommand{\href}[2]{#2}
\providecommand{\path}[1]{#1}
\providecommand{\DOIprefix}{doi:}
\providecommand{\ArXivprefix}{arXiv:}
\providecommand{\URLprefix}{URL: }
\providecommand{\Pubmedprefix}{pmid:}
\providecommand{\doi}[1]{\href{http://dx.doi.org/#1}{\path{#1}}}
\providecommand{\Pubmed}[1]{\href{pmid:#1}{\path{#1}}}
\providecommand{\bibinfo}[2]{#2}
\ifx\xfnm\relax \def\xfnm[#1]{\unskip,\space#1}\fi
\bibitem[{Beutel et~al.(2018)Beutel, Covington, Jain, Xu, Li, Gatto and Chi}]{beutel2018latent}
\bibinfo{author}{Beutel, A.}, \bibinfo{author}{Covington, P.}, \bibinfo{author}{Jain, S.}, \bibinfo{author}{Xu, C.}, \bibinfo{author}{Li, J.}, \bibinfo{author}{Gatto, V.}, \bibinfo{author}{Chi, E.H.}, \bibinfo{year}{2018}.
\newblock \bibinfo{title}{Latent cross: Making use of context in recurrent recommender systems}, in: \bibinfo{booktitle}{Proceedings of the eleventh ACM international conference on web search and data mining}, pp. \bibinfo{pages}{46--54}.
\bibitem[{Brown et~al.(2020)Brown, Mann, Ryder, Subbiah, Kaplan, Dhariwal, Neelakantan, Shyam, Sastry, Askell et~al.}]{brown2020language}
\bibinfo{author}{Brown, T.}, \bibinfo{author}{Mann, B.}, \bibinfo{author}{Ryder, N.}, \bibinfo{author}{Subbiah, M.}, \bibinfo{author}{Kaplan, J.D.}, \bibinfo{author}{Dhariwal, P.}, \bibinfo{author}{Neelakantan, A.}, \bibinfo{author}{Shyam, P.}, \bibinfo{author}{Sastry, G.}, \bibinfo{author}{Askell, A.}, et~al., \bibinfo{year}{2020}.
\newblock \bibinfo{title}{Language models are few-shot learners}.
\newblock \bibinfo{journal}{Advances in neural information processing systems} \bibinfo{volume}{33}, \bibinfo{pages}{1877--1901}.
\bibitem[{Catherine and Cohen(2017)}]{catherine2017transnets}
\bibinfo{author}{Catherine, R.}, \bibinfo{author}{Cohen, W.}, \bibinfo{year}{2017}.
\newblock \bibinfo{title}{Transnets: Learning to transform for recommendation}, in: \bibinfo{booktitle}{Proceedings of the eleventh ACM conference on recommender systems}, pp. \bibinfo{pages}{288--296}.
\bibitem[{Chen et~al.(2018)Chen, Xu, Zhang, Tang, Cao, Qin and Zha}]{chen2018sequential}
\bibinfo{author}{Chen, X.}, \bibinfo{author}{Xu, H.}, \bibinfo{author}{Zhang, Y.}, \bibinfo{author}{Tang, J.}, \bibinfo{author}{Cao, Y.}, \bibinfo{author}{Qin, Z.}, \bibinfo{author}{Zha, H.}, \bibinfo{year}{2018}.
\newblock \bibinfo{title}{Sequential recommendation with user memory networks}, in: \bibinfo{booktitle}{Proceedings of the eleventh ACM international conference on web search and data mining}, pp. \bibinfo{pages}{108--116}.
\bibitem[{Du et~al.(2023)Du, Yuan, Zhao, Fang, Liu, Liu, Sheng and Zhou}]{du2023contrastive}
\bibinfo{author}{Du, X.}, \bibinfo{author}{Yuan, H.}, \bibinfo{author}{Zhao, P.}, \bibinfo{author}{Fang, J.}, \bibinfo{author}{Liu, G.}, \bibinfo{author}{Liu, Y.}, \bibinfo{author}{Sheng, V.S.}, \bibinfo{author}{Zhou, X.}, \bibinfo{year}{2023}.
\newblock \bibinfo{title}{Contrastive enhanced slide filter mixer for sequential recommendation}, in: \bibinfo{booktitle}{2023 IEEE 39th International Conference on Data Engineering (ICDE)}, \bibinfo{organization}{IEEE}. pp. \bibinfo{pages}{2673--2685}.
\bibitem[{Ethayarajh(2019)}]{ethayarajh2019contextual}
\bibinfo{author}{Ethayarajh, K.}, \bibinfo{year}{2019}.
\newblock \bibinfo{title}{How contextual are contextualized word representations? comparing the geometry of bert, elmo, and gpt-2 embeddings}, in: \bibinfo{booktitle}{Proceedings of the 2019 Conference on Empirical Methods in Natural Language Processing and the 9th International Joint Conference on Natural Language Processing (EMNLP-IJCNLP)}, \bibinfo{organization}{Association for Computational Linguistics}.
\bibitem[{Friedman et~al.(2023)Friedman, Ahuja, Allen, Tan, Sidahmed, Long, Xie, Schubiner, Patel, Lara et~al.}]{friedman2023leveraging}
\bibinfo{author}{Friedman, L.}, \bibinfo{author}{Ahuja, S.}, \bibinfo{author}{Allen, D.}, \bibinfo{author}{Tan, T.}, \bibinfo{author}{Sidahmed, H.}, \bibinfo{author}{Long, C.}, \bibinfo{author}{Xie, J.}, \bibinfo{author}{Schubiner, G.}, \bibinfo{author}{Patel, A.}, \bibinfo{author}{Lara, H.}, et~al., \bibinfo{year}{2023}.
\newblock \bibinfo{title}{Leveraging large language models in conversational recommender systems}.
\newblock \bibinfo{journal}{arXiv preprint arXiv:2305.07961} .
\bibitem[{Gao et~al.(2023)Gao, Sheng, Xiang, Xiong, Wang and Zhang}]{gao2023chat}
\bibinfo{author}{Gao, Y.}, \bibinfo{author}{Sheng, T.}, \bibinfo{author}{Xiang, Y.}, \bibinfo{author}{Xiong, Y.}, \bibinfo{author}{Wang, H.}, \bibinfo{author}{Zhang, J.}, \bibinfo{year}{2023}.
\newblock \bibinfo{title}{Chat-rec: Towards interactive and explainable llms-augmented recommender system}.
\newblock \bibinfo{journal}{arXiv preprint arXiv:2303.14524} .
\bibitem[{Geng et~al.(2022)Geng, Liu, Fu, Ge and Zhang}]{geng2022recommendation}
\bibinfo{author}{Geng, S.}, \bibinfo{author}{Liu, S.}, \bibinfo{author}{Fu, Z.}, \bibinfo{author}{Ge, Y.}, \bibinfo{author}{Zhang, Y.}, \bibinfo{year}{2022}.
\newblock \bibinfo{title}{Recommendation as language processing (rlp): A unified pretrain, personalized prompt \& predict paradigm (p5)}, in: \bibinfo{booktitle}{Proceedings of the 16th ACM Conference on Recommender Systems}, pp. \bibinfo{pages}{299--315}.
\bibitem[{Hidasi and Karatzoglou(2018)}]{hidasi2018recurrent}
\bibinfo{author}{Hidasi, B.}, \bibinfo{author}{Karatzoglou, A.}, \bibinfo{year}{2018}.
\newblock \bibinfo{title}{Recurrent neural networks with top-k gains for session-based recommendations}, in: \bibinfo{booktitle}{Proceedings of the 27th ACM international conference on information and knowledge management}, pp. \bibinfo{pages}{843--852}.
\bibitem[{Hidasi et~al.(2015)Hidasi, Karatzoglou, Baltrunas and Tikk}]{hidasi2015session}
\bibinfo{author}{Hidasi, B.}, \bibinfo{author}{Karatzoglou, A.}, \bibinfo{author}{Baltrunas, L.}, \bibinfo{author}{Tikk, D.}, \bibinfo{year}{2015}.
\newblock \bibinfo{title}{Session-based recommendations with recurrent neural networks}.
\newblock \bibinfo{journal}{arXiv preprint arXiv:1511.06939} .
\bibitem[{Hou et~al.(2022)Hou, Mu, Zhao, Li, Ding and Wen}]{hou2022towards}
\bibinfo{author}{Hou, Y.}, \bibinfo{author}{Mu, S.}, \bibinfo{author}{Zhao, W.X.}, \bibinfo{author}{Li, Y.}, \bibinfo{author}{Ding, B.}, \bibinfo{author}{Wen, J.R.}, \bibinfo{year}{2022}.
\newblock \bibinfo{title}{Towards universal sequence representation learning for recommender systems}, in: \bibinfo{booktitle}{Proceedings of the 28th ACM SIGKDD Conference on Knowledge Discovery and Data Mining}, pp. \bibinfo{pages}{585--593}.
\bibitem[{Jing and Smola(2017)}]{jing2017neural}
\bibinfo{author}{Jing, H.}, \bibinfo{author}{Smola, A.J.}, \bibinfo{year}{2017}.
\newblock \bibinfo{title}{Neural survival recommender}, in: \bibinfo{booktitle}{Proceedings of the Tenth ACM International Conference on Web Search and Data Mining}, pp. \bibinfo{pages}{515--524}.
\bibitem[{Kang and McAuley(2018)}]{kang2018self}
\bibinfo{author}{Kang, W.C.}, \bibinfo{author}{McAuley, J.}, \bibinfo{year}{2018}.
\newblock \bibinfo{title}{Self-attentive sequential recommendation}, in: \bibinfo{booktitle}{2018 IEEE international conference on data mining (ICDM)}, \bibinfo{organization}{IEEE}. pp. \bibinfo{pages}{197--206}.
\bibitem[{Kim et~al.(2016)Kim, Park, Oh, Lee and Yu}]{kim2016convolutional}
\bibinfo{author}{Kim, D.}, \bibinfo{author}{Park, C.}, \bibinfo{author}{Oh, J.}, \bibinfo{author}{Lee, S.}, \bibinfo{author}{Yu, H.}, \bibinfo{year}{2016}.
\newblock \bibinfo{title}{Convolutional matrix factorization for document context-aware recommendation}, in: \bibinfo{booktitle}{Proceedings of the 10th ACM conference on recommender systems}, pp. \bibinfo{pages}{233--240}.
\bibitem[{Li et~al.(2020)Li, Zhou, He, Wang, Yang and Li}]{li2020sentence}
\bibinfo{author}{Li, B.}, \bibinfo{author}{Zhou, H.}, \bibinfo{author}{He, J.}, \bibinfo{author}{Wang, M.}, \bibinfo{author}{Yang, Y.}, \bibinfo{author}{Li, L.}, \bibinfo{year}{2020}.
\newblock \bibinfo{title}{On the sentence embeddings from pre-trained language models}, in: \bibinfo{booktitle}{Proceedings of the 2020 Conference on Empirical Methods in Natural Language Processing (EMNLP)}, pp. \bibinfo{pages}{9119--9130}.
\bibitem[{Li et~al.(2023a)Li, Wang, Li, Fu, Shen, Shang and McAuley}]{li2023text}
\bibinfo{author}{Li, J.}, \bibinfo{author}{Wang, M.}, \bibinfo{author}{Li, J.}, \bibinfo{author}{Fu, J.}, \bibinfo{author}{Shen, X.}, \bibinfo{author}{Shang, J.}, \bibinfo{author}{McAuley, J.}, \bibinfo{year}{2023}a.
\newblock \bibinfo{title}{Text is all you need: Learning language representations for sequential recommendation}, in: \bibinfo{booktitle}{Proceedings of the 29th ACM SIGKDD Conference on Knowledge Discovery and Data Mining}, pp. \bibinfo{pages}{1258--1267}.
\bibitem[{Li et~al.(2023b)Li, Zhang, Wang, Xiong, Lu and Medioni}]{li2023gpt4rec}
\bibinfo{author}{Li, J.}, \bibinfo{author}{Zhang, W.}, \bibinfo{author}{Wang, T.}, \bibinfo{author}{Xiong, G.}, \bibinfo{author}{Lu, A.}, \bibinfo{author}{Medioni, G.}, \bibinfo{year}{2023}b.
\newblock \bibinfo{title}{Gpt4rec: A generative framework for personalized recommendation and user interests interpretation}.
\newblock \bibinfo{journal}{arXiv preprint arXiv:2304.03879} .
\bibitem[{Li and She(2017)}]{li2017collaborative}
\bibinfo{author}{Li, X.}, \bibinfo{author}{She, J.}, \bibinfo{year}{2017}.
\newblock \bibinfo{title}{Collaborative variational autoencoder for recommender systems}, in: \bibinfo{booktitle}{Proceedings of the 23rd ACM SIGKDD international conference on knowledge discovery and data mining}, pp. \bibinfo{pages}{305--314}.
\bibitem[{Lipton et~al.(2015)Lipton, Berkowitz and Elkan}]{lipton2015critical}
\bibinfo{author}{Lipton, Z.C.}, \bibinfo{author}{Berkowitz, J.}, \bibinfo{author}{Elkan, C.}, \bibinfo{year}{2015}.
\newblock \bibinfo{title}{A critical review of recurrent neural networks for sequence learning}.
\newblock \bibinfo{journal}{arXiv preprint arXiv:1506.00019} .
\bibitem[{Liu et~al.(2023)Liu, Cai, Zhang, Zhao, Gao, Wang, Lv, Fan, Wang, He et~al.}]{liu2023linrec}
\bibinfo{author}{Liu, L.}, \bibinfo{author}{Cai, L.}, \bibinfo{author}{Zhang, C.}, \bibinfo{author}{Zhao, X.}, \bibinfo{author}{Gao, J.}, \bibinfo{author}{Wang, W.}, \bibinfo{author}{Lv, Y.}, \bibinfo{author}{Fan, W.}, \bibinfo{author}{Wang, Y.}, \bibinfo{author}{He, M.}, et~al., \bibinfo{year}{2023}.
\newblock \bibinfo{title}{Linrec: Linear attention mechanism for long-term sequential recommender systems}, in: \bibinfo{booktitle}{Proceedings of the 46th International ACM SIGIR Conference on Research and Development in Information Retrieval}, pp. \bibinfo{pages}{289--299}.
\bibitem[{Liu et~al.(2016)Liu, Wu, Wang, Li and Wang}]{liu2016context}
\bibinfo{author}{Liu, Q.}, \bibinfo{author}{Wu, S.}, \bibinfo{author}{Wang, D.}, \bibinfo{author}{Li, Z.}, \bibinfo{author}{Wang, L.}, \bibinfo{year}{2016}.
\newblock \bibinfo{title}{Context-aware sequential recommendation}, in: \bibinfo{booktitle}{2016 IEEE 16th International Conference on Data Mining (ICDM)}, \bibinfo{organization}{IEEE}. pp. \bibinfo{pages}{1053--1058}.
\bibitem[{Van~der Maaten and Hinton(2008)}]{van2008visualizing}
\bibinfo{author}{Van~der Maaten, L.}, \bibinfo{author}{Hinton, G.}, \bibinfo{year}{2008}.
\newblock \bibinfo{title}{Visualizing data using t-sne.}
\newblock \bibinfo{journal}{Journal of machine learning research} \bibinfo{volume}{9}.
\bibitem[{Ni et~al.(2019)Ni, Li and McAuley}]{ni2019justifying}
\bibinfo{author}{Ni, J.}, \bibinfo{author}{Li, J.}, \bibinfo{author}{McAuley, J.}, \bibinfo{year}{2019}.
\newblock \bibinfo{title}{Justifying recommendations using distantly-labeled reviews and fine-grained aspects}, in: \bibinfo{booktitle}{Proceedings of the 2019 conference on empirical methods in natural language processing and the 9th international joint conference on natural language processing (EMNLP-IJCNLP)}, pp. \bibinfo{pages}{188--197}.
\bibitem[{Radford et~al.(2019)Radford, Wu, Child, Luan, Amodei, Sutskever et~al.}]{radford2019language}
\bibinfo{author}{Radford, A.}, \bibinfo{author}{Wu, J.}, \bibinfo{author}{Child, R.}, \bibinfo{author}{Luan, D.}, \bibinfo{author}{Amodei, D.}, \bibinfo{author}{Sutskever, I.}, et~al., \bibinfo{year}{2019}.
\newblock \bibinfo{title}{Language models are unsupervised multitask learners}.
\newblock \bibinfo{journal}{OpenAI blog} \bibinfo{volume}{1}, \bibinfo{pages}{9}.
\bibitem[{Rendle et~al.(2010)Rendle, Freudenthaler and Schmidt-Thieme}]{rendle2010factorizing}
\bibinfo{author}{Rendle, S.}, \bibinfo{author}{Freudenthaler, C.}, \bibinfo{author}{Schmidt-Thieme, L.}, \bibinfo{year}{2010}.
\newblock \bibinfo{title}{Factorizing personalized markov chains for next-basket recommendation}, in: \bibinfo{booktitle}{Proceedings of the 19th international conference on World wide web}, pp. \bibinfo{pages}{811--820}.
\bibitem[{Su and Khoshgoftaar(2009)}]{su2009survey}
\bibinfo{author}{Su, X.}, \bibinfo{author}{Khoshgoftaar, T.M.}, \bibinfo{year}{2009}.
\newblock \bibinfo{title}{A survey of collaborative filtering techniques}.
\newblock \bibinfo{journal}{Advances in artificial intelligence} \bibinfo{volume}{2009}.
\bibitem[{Sun et~al.(2019)Sun, Liu, Wu, Pei, Lin, Ou and Jiang}]{sun2019bert4rec}
\bibinfo{author}{Sun, F.}, \bibinfo{author}{Liu, J.}, \bibinfo{author}{Wu, J.}, \bibinfo{author}{Pei, C.}, \bibinfo{author}{Lin, X.}, \bibinfo{author}{Ou, W.}, \bibinfo{author}{Jiang, P.}, \bibinfo{year}{2019}.
\newblock \bibinfo{title}{Bert4rec: Sequential recommendation with bidirectional encoder representations from transformer}, in: \bibinfo{booktitle}{Proceedings of the 28th ACM international conference on information and knowledge management}, pp. \bibinfo{pages}{1441--1450}.
\bibitem[{Wang et~al.(2022)Wang, Yu, Ma, Zhang, Chen, Liu and Ma}]{wang2022towards}
\bibinfo{author}{Wang, C.}, \bibinfo{author}{Yu, Y.}, \bibinfo{author}{Ma, W.}, \bibinfo{author}{Zhang, M.}, \bibinfo{author}{Chen, C.}, \bibinfo{author}{Liu, Y.}, \bibinfo{author}{Ma, S.}, \bibinfo{year}{2022}.
\newblock \bibinfo{title}{Towards representation alignment and uniformity in collaborative filtering}, in: \bibinfo{booktitle}{Proceedings of the 28th ACM SIGKDD Conference on Knowledge Discovery and Data Mining}, pp. \bibinfo{pages}{1816--1825}.
\bibitem[{Wang et~al.(2019a)Wang, Huang, Huang, Hu, Wang and Gu}]{wang2019improving}
\bibinfo{author}{Wang, L.}, \bibinfo{author}{Huang, J.}, \bibinfo{author}{Huang, K.}, \bibinfo{author}{Hu, Z.}, \bibinfo{author}{Wang, G.}, \bibinfo{author}{Gu, Q.}, \bibinfo{year}{2019}a.
\newblock \bibinfo{title}{Improving neural language generation with spectrum control}, in: \bibinfo{booktitle}{International Conference on Learning Representations}.
\bibitem[{Wang et~al.(2019b)Wang, Hu, Wang, Cao, Sheng and Orgun}]{wang2019sequential}
\bibinfo{author}{Wang, S.}, \bibinfo{author}{Hu, L.}, \bibinfo{author}{Wang, Y.}, \bibinfo{author}{Cao, L.}, \bibinfo{author}{Sheng, Q.Z.}, \bibinfo{author}{Orgun, M.}, \bibinfo{year}{2019}b.
\newblock \bibinfo{title}{Sequential recommender systems: challenges, progress and prospects}, in: \bibinfo{booktitle}{28th International Joint Conference on Artificial Intelligence, IJCAI 2019}, \bibinfo{organization}{International Joint Conferences on Artificial Intelligence}. pp. \bibinfo{pages}{6332--6338}.
\bibitem[{Wang and Isola(2020)}]{wang2020understanding}
\bibinfo{author}{Wang, T.}, \bibinfo{author}{Isola, P.}, \bibinfo{year}{2020}.
\newblock \bibinfo{title}{Understanding contrastive representation learning through alignment and uniformity on the hypersphere}, in: \bibinfo{booktitle}{International Conference on Machine Learning}, \bibinfo{organization}{PMLR}. pp. \bibinfo{pages}{9929--9939}.
\bibitem[{Wang et~al.(2024)Wang, Yu, Gao, Yin, Cui and Sadiq}]{wang2024unveiling}
\bibinfo{author}{Wang, Z.}, \bibinfo{author}{Yu, J.}, \bibinfo{author}{Gao, M.}, \bibinfo{author}{Yin, H.}, \bibinfo{author}{Cui, B.}, \bibinfo{author}{Sadiq, S.}, \bibinfo{year}{2024}.
\newblock \bibinfo{title}{Unveiling vulnerabilities of contrastive recommender systems to poisoning attacks}, in: \bibinfo{booktitle}{Proceedings of the 30th ACM SIGKDD Conference on Knowledge Discovery and Data Mining}, pp. \bibinfo{pages}{3311--3322}.
\bibitem[{Wu et~al.(2019)Wu, Wu, An, Huang, Huang and Xie}]{wu2019npa}
\bibinfo{author}{Wu, C.}, \bibinfo{author}{Wu, F.}, \bibinfo{author}{An, M.}, \bibinfo{author}{Huang, J.}, \bibinfo{author}{Huang, Y.}, \bibinfo{author}{Xie, X.}, \bibinfo{year}{2019}.
\newblock \bibinfo{title}{Npa: neural news recommendation with personalized attention}, in: \bibinfo{booktitle}{Proceedings of the 25th ACM SIGKDD international conference on knowledge discovery \& data mining}, pp. \bibinfo{pages}{2576--2584}.
\bibitem[{Wu et~al.(2017)Wu, Ahmed, Beutel and Smola}]{wu2017joint}
\bibinfo{author}{Wu, C.Y.}, \bibinfo{author}{Ahmed, A.}, \bibinfo{author}{Beutel, A.}, \bibinfo{author}{Smola, A.J.}, \bibinfo{year}{2017}.
\newblock \bibinfo{title}{Joint training of ratings and reviews with recurrent recommender networks}.
\newblock \URLprefix \url{https://openreview.net/forum?id=HksioDcxl}.
\bibitem[{Wu et~al.(2022)Wu, He, Wang, Zhang and Wang}]{wu2022survey}
\bibinfo{author}{Wu, L.}, \bibinfo{author}{He, X.}, \bibinfo{author}{Wang, X.}, \bibinfo{author}{Zhang, K.}, \bibinfo{author}{Wang, M.}, \bibinfo{year}{2022}.
\newblock \bibinfo{title}{A survey on accuracy-oriented neural recommendation: From collaborative filtering to information-rich recommendation}.
\newblock \bibinfo{journal}{IEEE Transactions on Knowledge and Data Engineering} \bibinfo{volume}{35}, \bibinfo{pages}{4425--4445}.
\bibitem[{Xu et~al.(2019)Xu, Zhao, Liu, Sheng, Xu, Zhuang, Fang and Zhou}]{xu2019graph}
\bibinfo{author}{Xu, C.}, \bibinfo{author}{Zhao, P.}, \bibinfo{author}{Liu, Y.}, \bibinfo{author}{Sheng, V.S.}, \bibinfo{author}{Xu, J.}, \bibinfo{author}{Zhuang, F.}, \bibinfo{author}{Fang, J.}, \bibinfo{author}{Zhou, X.}, \bibinfo{year}{2019}.
\newblock \bibinfo{title}{Graph contextualized self-attention network for session-based recommendation.}, in: \bibinfo{booktitle}{IJCAI}, pp. \bibinfo{pages}{3940--3946}.
\bibitem[{Yang et~al.(2023)Yang, Huang, Xia, Huang, Luo and Lin}]{yang2023debiased}
\bibinfo{author}{Yang, Y.}, \bibinfo{author}{Huang, C.}, \bibinfo{author}{Xia, L.}, \bibinfo{author}{Huang, C.}, \bibinfo{author}{Luo, D.}, \bibinfo{author}{Lin, K.}, \bibinfo{year}{2023}.
\newblock \bibinfo{title}{Debiased contrastive learning for sequential recommendation}, in: \bibinfo{booktitle}{Proceedings of the ACM Web Conference 2023}, pp. \bibinfo{pages}{1063--1073}.
\bibitem[{Yu et~al.(2022)Yu, Yin, Xia, Chen, Cui and Nguyen}]{yu2022graph}
\bibinfo{author}{Yu, J.}, \bibinfo{author}{Yin, H.}, \bibinfo{author}{Xia, X.}, \bibinfo{author}{Chen, T.}, \bibinfo{author}{Cui, L.}, \bibinfo{author}{Nguyen, Q.V.H.}, \bibinfo{year}{2022}.
\newblock \bibinfo{title}{Are graph augmentations necessary? simple graph contrastive learning for recommendation}, in: \bibinfo{booktitle}{Proceedings of the 45th international ACM SIGIR conference on research and development in information retrieval}, pp. \bibinfo{pages}{1294--1303}.
\bibitem[{Yuan et~al.(2019)Yuan, Karatzoglou, Arapakis, Jose and He}]{yuan2019simple}
\bibinfo{author}{Yuan, F.}, \bibinfo{author}{Karatzoglou, A.}, \bibinfo{author}{Arapakis, I.}, \bibinfo{author}{Jose, J.M.}, \bibinfo{author}{He, X.}, \bibinfo{year}{2019}.
\newblock \bibinfo{title}{A simple convolutional generative network for next item recommendation}, in: \bibinfo{booktitle}{Proceedings of the twelfth ACM international conference on web search and data mining}, pp. \bibinfo{pages}{582--590}.
\bibitem[{Yuan et~al.(2023)Yuan, Yuan, Song, Li, Fu, Yang, Pan and Ni}]{yuan2023go}
\bibinfo{author}{Yuan, Z.}, \bibinfo{author}{Yuan, F.}, \bibinfo{author}{Song, Y.}, \bibinfo{author}{Li, Y.}, \bibinfo{author}{Fu, J.}, \bibinfo{author}{Yang, F.}, \bibinfo{author}{Pan, Y.}, \bibinfo{author}{Ni, Y.}, \bibinfo{year}{2023}.
\newblock \bibinfo{title}{Where to go next for recommender systems? id-vs. modality-based recommender models revisited}, in: \bibinfo{booktitle}{Proceedings of the 46th International ACM SIGIR Conference on Research and Development in Information Retrieval}, pp. \bibinfo{pages}{2639--2649}.
\bibitem[{Zhang et~al.(2024a)Zhang, Zhou, Zeng and Shen}]{zhang2024id_embeddings}
\bibinfo{author}{Zhang, L.}, \bibinfo{author}{Zhou, X.}, \bibinfo{author}{Zeng, Z.}, \bibinfo{author}{Shen, Z.}, \bibinfo{year}{2024}a.
\newblock \bibinfo{title}{Are id embeddings necessary? whitening pre-trained text embeddings for effective sequential recommendation}, in: \bibinfo{booktitle}{2024 IEEE 40th International Conference on Data Engineering (ICDE)}, \bibinfo{publisher}{IEEE}.
\bibitem[{Zhang et~al.(2022)Zhang, Wu, Yu, Liu and Wang}]{zhang2022dynamic}
\bibinfo{author}{Zhang, M.}, \bibinfo{author}{Wu, S.}, \bibinfo{author}{Yu, X.}, \bibinfo{author}{Liu, Q.}, \bibinfo{author}{Wang, L.}, \bibinfo{year}{2022}.
\newblock \bibinfo{title}{Dynamic graph neural networks for sequential recommendation}.
\newblock \bibinfo{journal}{IEEE Transactions on Knowledge and Data Engineering} \bibinfo{volume}{35}, \bibinfo{pages}{4741--4753}.
\bibitem[{Zhang et~al.(2019)Zhang, Yao, Sun and Tay}]{zhang2019deep}
\bibinfo{author}{Zhang, S.}, \bibinfo{author}{Yao, L.}, \bibinfo{author}{Sun, A.}, \bibinfo{author}{Tay, Y.}, \bibinfo{year}{2019}.
\newblock \bibinfo{title}{Deep learning based recommender system: A survey and new perspectives}.
\newblock \bibinfo{journal}{ACM computing surveys (CSUR)} \bibinfo{volume}{52}, \bibinfo{pages}{1--38}.
\bibitem[{Zhang et~al.(2024b)Zhang, Bao, Yan, Wang, Feng and He}]{zhang2024text}
\bibinfo{author}{Zhang, Y.}, \bibinfo{author}{Bao, K.}, \bibinfo{author}{Yan, M.}, \bibinfo{author}{Wang, W.}, \bibinfo{author}{Feng, F.}, \bibinfo{author}{He, X.}, \bibinfo{year}{2024}b.
\newblock \bibinfo{title}{Text-like encoding of collaborative information in large language models for recommendation}.
\newblock \bibinfo{journal}{arXiv preprint arXiv:2406.03210} .
\bibitem[{Zhang and Sabuncu(2018)}]{zhang2018generalized}
\bibinfo{author}{Zhang, Z.}, \bibinfo{author}{Sabuncu, M.}, \bibinfo{year}{2018}.
\newblock \bibinfo{title}{Generalized cross entropy loss for training deep neural networks with noisy labels}.
\newblock \bibinfo{journal}{Advances in neural information processing systems} \bibinfo{volume}{31}.
\bibitem[{Zheng et~al.(2017)Zheng, Noroozi and Yu}]{zheng2017joint}
\bibinfo{author}{Zheng, L.}, \bibinfo{author}{Noroozi, V.}, \bibinfo{author}{Yu, P.S.}, \bibinfo{year}{2017}.
\newblock \bibinfo{title}{Joint deep modeling of users and items using reviews for recommendation}, in: \bibinfo{booktitle}{Proceedings of the tenth ACM international conference on web search and data mining}, pp. \bibinfo{pages}{425--434}.
\bibitem[{Zheng et~al.(2024)Zheng, Chao, Qiu, Zhu and Xiong}]{zheng2024harnessing}
\bibinfo{author}{Zheng, Z.}, \bibinfo{author}{Chao, W.}, \bibinfo{author}{Qiu, Z.}, \bibinfo{author}{Zhu, H.}, \bibinfo{author}{Xiong, H.}, \bibinfo{year}{2024}.
\newblock \bibinfo{title}{Harnessing large language models for text-rich sequential recommendation}, in: \bibinfo{booktitle}{Proceedings of the ACM on Web Conference 2024}, pp. \bibinfo{pages}{3207--3216}.
\bibitem[{Zhou et~al.(2022)Zhou, Yu, Zhao and Wen}]{zhou2022filter}
\bibinfo{author}{Zhou, K.}, \bibinfo{author}{Yu, H.}, \bibinfo{author}{Zhao, W.X.}, \bibinfo{author}{Wen, J.R.}, \bibinfo{year}{2022}.
\newblock \bibinfo{title}{Filter-enhanced mlp is all you need for sequential recommendation}, in: \bibinfo{booktitle}{Proceedings of the ACM web conference 2022}, pp. \bibinfo{pages}{2388--2399}.
\bibitem[{Zhou et~al.(2023)Zhou, Gao, Xie, Ye, Hua, Kim, Wang and Kim}]{zhou2023equivariant}
\bibinfo{author}{Zhou, P.}, \bibinfo{author}{Gao, J.}, \bibinfo{author}{Xie, Y.}, \bibinfo{author}{Ye, Q.}, \bibinfo{author}{Hua, Y.}, \bibinfo{author}{Kim, J.}, \bibinfo{author}{Wang, S.}, \bibinfo{author}{Kim, S.}, \bibinfo{year}{2023}.
\newblock \bibinfo{title}{Equivariant contrastive learning for sequential recommendation}, in: \bibinfo{booktitle}{Proceedings of the 17th ACM Conference on Recommender Systems}, pp. \bibinfo{pages}{129--140}.
\bibitem[{Zhu et~al.(2019)Zhu, Zhou, Song, Tan and Guo}]{zhu2019dan}
\bibinfo{author}{Zhu, Q.}, \bibinfo{author}{Zhou, X.}, \bibinfo{author}{Song, Z.}, \bibinfo{author}{Tan, J.}, \bibinfo{author}{Guo, L.}, \bibinfo{year}{2019}.
\newblock \bibinfo{title}{Dan: Deep attention neural network for news recommendation}, in: \bibinfo{booktitle}{Proceedings of the AAAI Conference on Artificial Intelligence}, pp. \bibinfo{pages}{5973--5980}.
\bibitem[{Zhu et~al.(2024)Zhu, Li, Liu and Luo}]{zhu2024multi}
\bibinfo{author}{Zhu, X.}, \bibinfo{author}{Li, L.}, \bibinfo{author}{Liu, W.}, \bibinfo{author}{Luo, X.}, \bibinfo{year}{2024}.
\newblock \bibinfo{title}{Multi-level sequence denoising with cross-signal contrastive learning for sequential recommendation}.
\newblock \bibinfo{journal}{Neural Networks} \bibinfo{volume}{179}, \bibinfo{pages}{106480}.

\end{thebibliography}

\end{document}